\documentclass[reprint,superscriptaddress,amsmath,amssym]{revtex4-2}

\usepackage{bm}
\usepackage{float}
\usepackage{graphicx}
\usepackage[usenames,dvipsnames]{color}
\usepackage[normalem]{ulem}
\usepackage[svgnames]{xcolor}
\usepackage{bm}
\usepackage{multirow}
\usepackage{float}
\usepackage{titlesec}
\usepackage[utf8]{inputenc}

\begin{document}

\title{Phonon hydrodynamics in bulk insulators and semi-metals}
\author{Yo Machida}
\affiliation{Department of Physics, Gakushuin University, Tokyo 171-8588, Japan.}
\author{Valentina Martelli}
\affiliation{Laboratory for Quantum Matter under Extreme Conditions\\ Institute of Physics, University of São Paulo, 05508-090, São Paulo, Brazil}
\author{Alexandre Jaoui}
\affiliation{JEIP, USR 3573 CNRS, Coll\`ege de France, PSL Research University, 11, Place Marcelin Berthelot, 75231 Paris Cedex 05, France}
\affiliation{Laboratoire de Physique et d'\'Etude des Mat\'eriaux \\ (ESPCI - CNRS - Sorbonne Universit\'e), PSL Research University, 75005 Paris, France}

\author{Beno\^it Fauqu\'e}
\affiliation{JEIP, USR 3573 CNRS, Coll\`ege de France, PSL Research University, 11, Place Marcelin Berthelot, 75231 Paris Cedex 05, France}
\affiliation{Laboratoire de Physique et d'\'Etude des Mat\'eriaux \\ (ESPCI - CNRS - Sorbonne Universit\'e), PSL Research University, 75005 Paris, France}

\author{Kamran Behnia}
\affiliation{Laboratoire de Physique et d'\'Etude des Mat\'eriaux \\ (ESPCI - CNRS - Sorbonne Universit\'e), PSL Research University, 75005 Paris, France}

\date{\today}

\begin{abstract}
 Decades ago, Gurzhi proposed that if momentum-conserving collisions prevail among heat-carrying phonons in insulators and charge-carrying electrons in metals, hydrodynamic features will become detectable. In this paper, we will review the experimental evidence emerging in the last few years supporting this viewpoint and raising new questions. The focus of the paper will be bulk crystals without (or with a very dilute concentration of) mobile electrons and steady-state thermal transport. We also discuss the possible link between this field of investigation and other phenomena, such as the hybridization of phonon modes and the phonon thermal Hall effect.
\end{abstract}
\maketitle

\section{Introduction}
In his seminal 1968 paper, entitled ``Hydrodynamic effects in solids at low temperatures''\cite{gurzhi1968}, Gurzhi wrote: ``The phenomena of thermal conductivity of dielectric [i.e. insulators] and the electrical conductivity of metals have specific properties. In both cases, the total quasi-particle current turns out to be non-vanishing. It follows that when only normal collisions occur in the system, there could exist an undamped current in the absence of an external field which could sustain it.''  The adjective `normal' is employed here to specify those collisions (we will call them `Normal'), which are not of the Umklapp type, i.e. do not produce a change in quasimomentum larger than the width of the Brillouin zone. 

In the text quoted above, Gurzhi is proposing a thought experiment. For both electrons in metals and phonons in insulators, one is invited to wonder what happens if Umklapp collisions are turned off. He proposes that then if there were no viscosity, the motion would become perpetual. This insight opens a route towards exploration of the viscosity of these quasi-particles. 

\begin{figure*}[ht]
\begin{center}
\includegraphics[width=15cm]{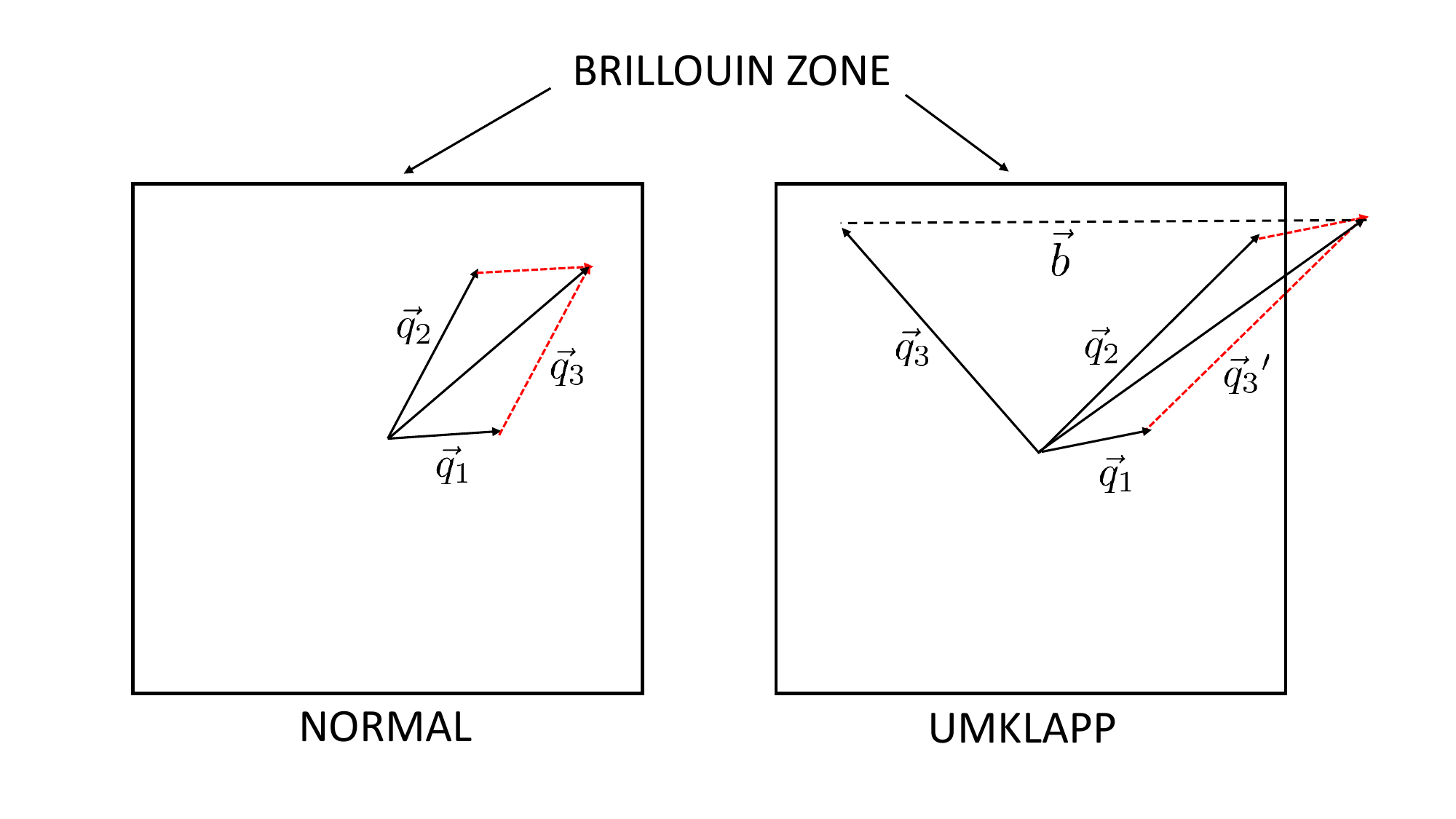}
\caption{\textbf{The contrast between Normal and Umklapp events during a three phonon process:}
In a Normal collision (left), the resultant phonon wave-vector remains inside the Brillouin zone. This is not the case in an Umklapp event (right), which leads to the loss of momentum  $\overrightarrow{b}$, corresponding to a unit vector of the reciprocal lattice (adapted from \cite{Vandersande}).}
\label{Fig-NU}
\end{center}
\end{figure*}

When two phonons collide, a third phonon is produced. If the wave-vector of the latter is larger than the size of the Brillouin zone, then the collision is called Umklapp (See Fig.\ref{Fig-NU}, which was adapted from Ref. \cite{Vandersande}). Such an event would degrade the momentum (and the energy) flow of phonons. In contrast, during a Normal collision, the wave-vector of the third phonon remains shorter than the width of the Brillouin zone and the flow is not degraded. The same is true for electrons, where because of the conservation of charge, the two initial electrons are replaced by two and not one. 

Now, there is a fundamental difference between electrons and phonons regarding the Umklapp collisions. The wave-vector of a thermally excited phonon $q_{ph}$ decreases with temperature. At a given temperature, it is $q_{ph}=\frac{k_BT}{\hbar v_s}$, where $v_s$ is the sound velocity, $k_B$ is the Boltzmann constant, and $\hbar$ is the reduced Planck constant. Therefore,  in order to perform Gurzhi's thought experiment for phonons, all you need [to get rid of Umklapp events] is to cool down your sample well below a temperature, where   $q_{ph}< G/2$, which corresponds to a fraction of the Debye temperature.  The situation is quite different for electrons, whose wavevector is set by the Fermi radius and does not change with temperature. A very small Fermi surface, located at the center of the Brillouin zone \cite{lin2015} is needed to avoid Umklapp collisions. 

The role of disorder is another factor which renders the observation of phonon hydrodynamics easier than its electronic counterpart. Since the wavelength of heat-carrying acoustic phonons increases with cooling, point defects become incapable of scattering them. In a single crystalline bulk insulator, the phonon mean free path becomes equal to the size of the sample at cryogenic temperatures, and the default transport regime becomes ballistic. Achieving this for electrons in metals is much more challenging.  

These fundamental differences may explain, at least partially, the historical fact that phonon hydrodynamics was explored before electron hydrodynamics. As early as 1974, two decades before the first recorded experimental investigation of electron hydrodynamics \cite{molenkamp1995}, Beck and co-workers \cite{beck1974} wrote a review paper on phonon hydrodynamics.  

Second sound refers to the wave-like propagation of temperature and entropy \cite{Donnelly2009}. It was first observed in superfluid helium by Peshkov in 1946 \cite{Peshkov1946, Peshkov1960}. In 1952, Dingle suggested that a density fluctuation can propagate as a second sound wave in a phonon gas \cite{Dingle_1952}. Guyer and Krumhansl \cite{Guyer1966} argued that the Poiseuille flow of phonons in a steady-state experiment is concomitant with the occurrence of phononic second sound, both requiring the same hierarchy of scattering times. Thanks to several  works \cite{Gurevich1967,Nielsen1969, Mezhov1980,Gurevich1986}, it was established that a hydrodynamic window emerges between the ballistic and diffusive transport regimes when momentum-conserving collisions between phonons become sufficiently abundant.

After a promising start, this field of research  went through hibernation in the last decades of the previous century. The renaissance of this field of research began in 2015 with theoretical studies of heat transport in graphene \cite{Cepellotti2015,Lee2015}.   The present paper summarizes the evidence gathered in the last few years by steady-sate transport measurements, confirming the existence of a hydrodynamic window for thermal transport by phonons. Our main focus will be a subset of such hydrodynamic phenomena associated with the Poiseuille regime observed in bulk crystals by steady-state heat transport studies.  For a recent review paper on phonon hydrodynamics, written from a theoretical perspective, and covering a broader range of topics, see \cite{Ghosh_2022}.

\section{Heat transport regimes}

\begin{figure*}[ht]
\begin{center}
\includegraphics[width=15cm]{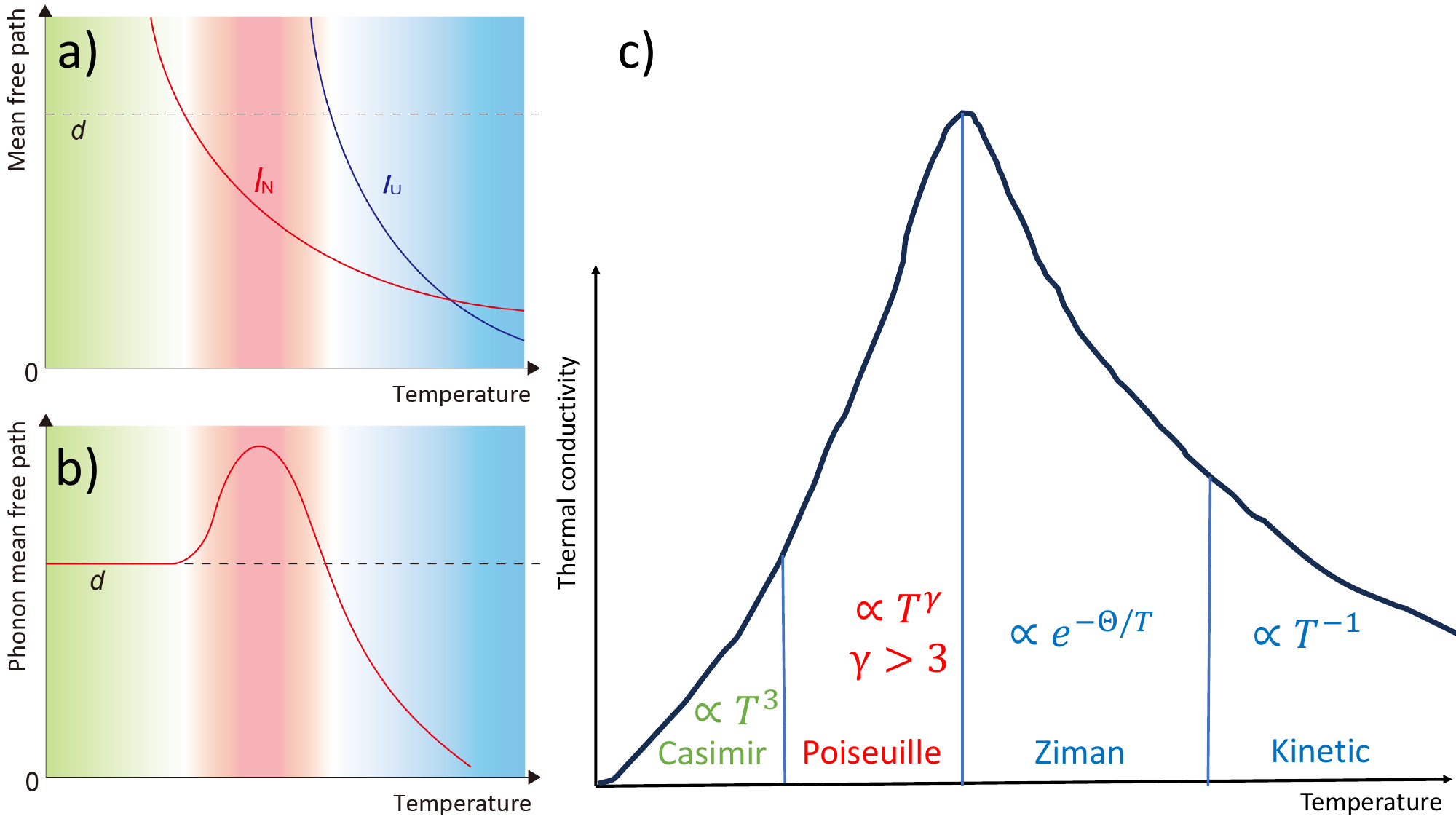}
\caption{\textbf{The four regimes of heat transport and the hydrodynamic window:}
a) The distance between two Normal collisions $l_N$ and two Umklapp collisions $l_U$ increases with cooling. But the increase in $l_N$ is slower.  Eventually, both exceed the sample thickness, $d$, shown by a horizontal dashed line. The sketch distinguishes between three regimes. In the ballistic regime (in green), the most frequent collisions suffered by a heat-carrying phonon are boundary collisions.  In the diffusive regime (in blue), Umklapp collisions are more frequent than collisions with boundary. In the hydrodynamic regime (in red), Boundary collisions are more frequent than Umklapp collisions, but less frequent than Normal collisions. b) A sketch of the evolution of the effective phonon mean free path with temperature. In the hydrodynamic regime, it exceeds the sample thickness. c) The four regimes of heat transport in a crystalline insulator (adapted from \cite{beck1974}). The Poiseuille regime is where thermal conductivity evolves faster than T$^3$.}
\label{Fig-regimes}
\end{center}
\end{figure*}

Figure \ref{Fig-regimes} presents a generic picture of the evolution of fundamental length scales and their consequences for thermal transport by phonons in an insulator. Near the Debye temperature and above, the phonon wave-vector is comparable to the width of the Brillouin zone, and most phonon-phonon collisions are Umklapp. Therefore: $l_N >l_U$; that is, the distance between two Umklapp collisions ($l_U$) is shorter than the one between two Normal collisions ($l_N$). As the temperature decreases and the phonon wave-vector shrinks, the frequency of Umklapp collisions decreases. Since the phonon density of states also decreases with cooling, Normal collisions rarefy too, albeit with a slower rate (Figure \ref{Fig-regimes}a). As a consequence of this inequality, an intermediate temperature window (in red) can pop up between the high-temperature diffusive (in green) and the low-temperature ballistic regimes (in blue). 
In the ballistic regime, the shortest length scale is the [temperature-independent] thickness of the sample (\textit{d}) and transport is dominated by boundary scattering. In the diffusive regime, transport is dominated by phonon-phonon scattering, which is mostly Umklapp. Thermal conductivity in the diffusive regime is intrinsic to the insulating solid, does not depend on the sample size and can be quantified nowadays by \textit{ab initio} theory starting from the computed phonon spectrum. It steadily \textit{decreases} with rising temperature. In contrast, thermal conductivity in the ballistic regime depends on the size of the crystal,  the velocity of sound and the heat capacity, and because of the latter, \textit{increases} with rising temperature. The hydrodynamic regime, which may arise between these two, is characterized by a specific hierarchy of the three length scales: $l_U>d>l_N$. It implies that compared to boundary scattering, Umklapp collisions are rare and Normal collisions are abundant. One remarkable consequence of the emergence of such a hydrodynamic regime is sketched in Figure \ref{Fig-regimes}b. Thanks to Normal collisions, the temperature dependence of the effective phonon mean-path becomes non-monotonous and its peak value is expected to be longer than the effective thickness of the sample. 

Based on the hierarchy of distinct scattering times and mean free paths,  Beck, Meier and Thellung \cite{beck1974} identified four regimes of heat transport in a generic insulator (See also \cite{Cepellotti2015} and \cite{Ghosh_2022}). They are illustrated, together, with the expected temperature dependence of thermal conductivity in each regime, in Figure \ref{Fig-regimes}c.  

In the high-temperature regime, called `kinetic', Umklapp phonon-phonon scattering dominates and $l_U$ is the shortest length scale. The temperature is high enough for thermal excitation of almost all phonon population. The specific heat tends to saturate to its universal Dulong-Petit value of 3k$_B$ per atom and the thermal conductivity tends to follow the inverse of temperature. It has been recently noted that the phonon-phonon scattering time in this regime approaches but never falls below the so-called Planckian time \cite{martelli2018,Behnia_2019}. Mousatov and Hartnoll \cite{Mousatov2020} have argued that this bound is due to a bound on the sound velocity. Because of quantum mechanics, it cannot fall below a velocity set by the melting temperature of the crystal, the interatomic spacing and $\hbar$. 

Cooling the crystal, one enters the so-called 'Ziman' regime \cite{beck1974}. In this temperature range, phonons with a large wave-vector are absent, and therefore, Umklapp collisions become rare. As a consequence, $l_N<l_U$. Both $l_N$ and $l_U$  remain shorter than the sample size, $d$. In this regime, thermal conductivity increases faster than linear with decreasing temperature and is expected to become eventually exponential, following $\propto \exp(-\frac{\Theta}{T})$. However, a clear exponential behavior has only been seen in a handful of solids \cite{berman1976thermal}. 

Cooling further, leads to a peak in thermal conductivity. At this peak temperature, the decrease in phonon population and the increase in the phonon scattering time, both induced by cooling cancel each other. Below this peak, the temperature dependence of $\kappa$ is mainly set by the temperature dependence of specific heat, which, asymptotically, is $\propto T^3$. The ballistic regime, also known as the Casimir regime \cite{beck1974}, referring to an early and influential note by Casimir \cite{CASIMIR1938495},  occurs at the lowest temperatures. Here,  $l_N$ and $l_U$ become longer than the sample size, which sets the ultimate mean free path of the heat carriers. Thermal conductivity in single crystals becomes size dependent \cite{Thacher1967} and follows a $\propto T^3$ temperature dependence.

The Poiseuille regime, which is the focus of the present paper is expected to emerge between the ballistic regime and the peak. In the Poiseuille regime, one has $l_N < d< l_U$. Now, most collision events are Normal  and the frequency of Umklapp collisions is negligible. Therefore phonons can exchange momentum frequently before colliding with the boundaries. In this case, phonons closer to the boundary would have  a drift velocity lower than those further from the boundaries. In this temperature range, the thermal conductivity is expected to have a temperature dependence faster than the specific heat. This implies that warming the sample leads to an increase in the effective mean free path, despite an increase in the collision rate.
\begin{figure*}[ht]
\begin{center}
\includegraphics[width=17cm]{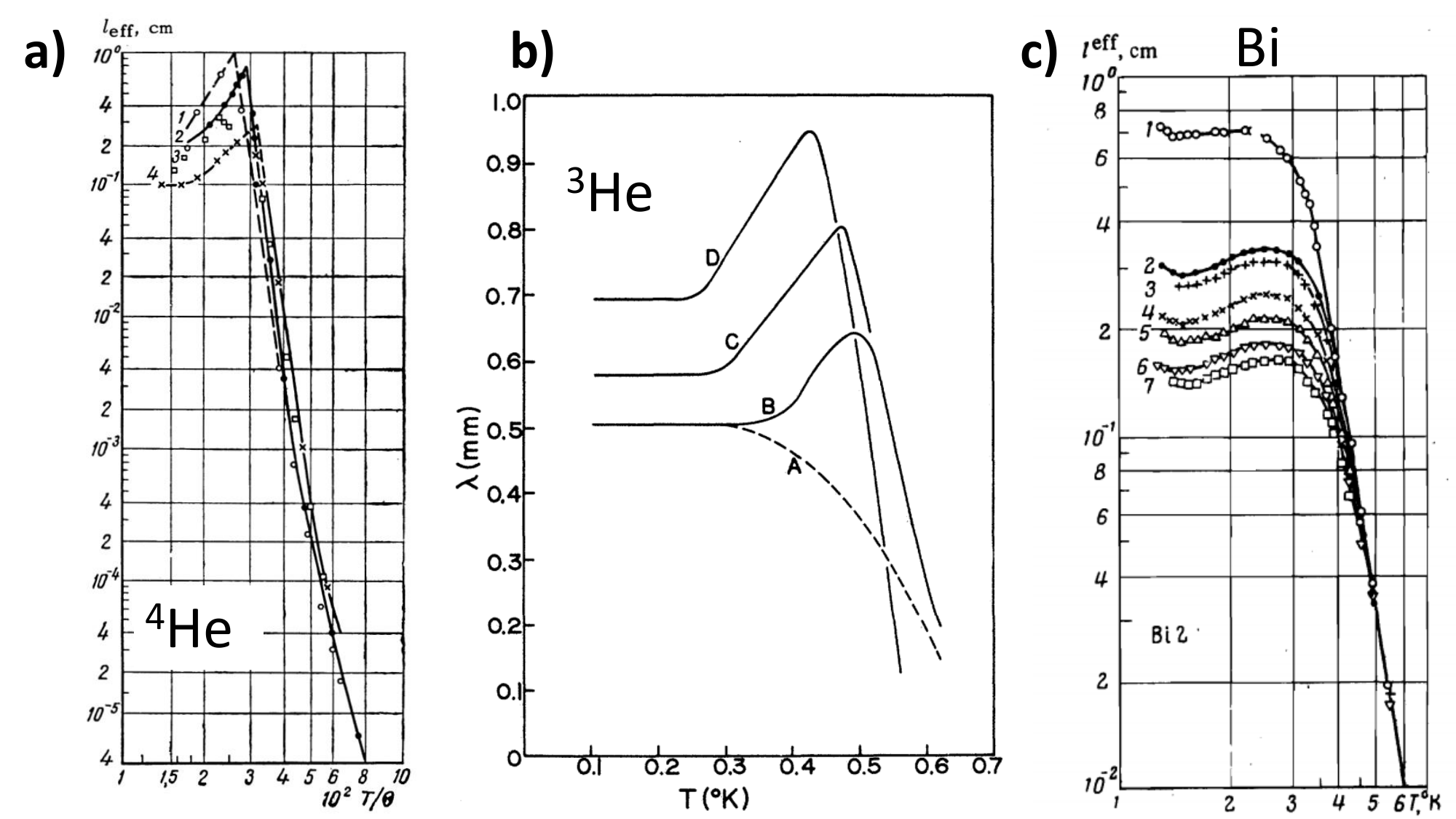}
\caption{\textbf{Early observations:} a) The variation of the effective phonon mean free path with temperature (normalized by the Debye temperature) in different crystals of $^4$He grown under different pressures from 60 (sample 1)  to 185 (sample 4) bars. Note the presence of a peak [From ref \cite{mezhov1966}].  b) Temperature dependence of the effective phonon mean free path in Body Centered Cubic crystals of $^3$He grown in pressures ranging from 56 (sample D) to 96  (sample A) bars.  There is a local peak [From ref \cite{thomlinson1969}]. c) Temperature dependence of the effective phonon mean free path in Bi crystals with different sizes. There is  clearly a local peak and a size dependence. However, the size dependence at the peak and below the peak are not very different [From ref.\cite{kopylov1973}]. }
\label{Fig-old}
\end{center}
\end{figure*}

\section{Thermal conductivity in the ideal Poiseuille regime}

With cooling, the collision rate between phonons is reduced and both l$_U$ and l$_N$ increase. In the simplest picture of phonon transport, l$_U$ rises exponentially and l$_N$ follows a power law. The expected temperature dependence of $\kappa$ in the Poiseuille regime  depends on the exponent of this power law. A simple argument would lead to l$_N \propto T^{-5}$, reminiscent of the temperature dependence of the e-ph scattering time for temperatures well below the Debye temperature \cite{ziman2001electrons}. The similarity is not accidental, because the origin is the same. The volume of Debye sphere, which sets the phonon population is $\propto T^{3}$. On top of this, the collisions cross section, which is $\propto q^{2}$, will add another $\propto T^{2}$ factor, leading to a $\propto T^{5}$ phonon-phonon Normal scattering rate. 

Consider that thermal conductivity can be expressed as $\kappa=1/3C_Vv_s\ell_{eff}$. The effective mean free path $\ell_{eff}$ is expected to be proportional to the cross section of the conductor and inversely proportional to the distance between Normal collisions:  $\ell_{eff}=\frac{d^2}{l_N}$. Therefore, in an ideal Poiseuille regime one expects:
\begin{equation}
    \kappa=1/3C_Vv_s\frac{d^2}{l_N}
\end{equation}

Thus, one expects two scaling relations for $\kappa$. A temperature dependence as strong as $\propto T^8$ and a super-ballistic size dependence $\propto d^2$ \cite{beck1974}. 

\maketitle\section{Experimental detection of the Poiseuille regime}
Experiments have found signatures of hydrodynamic heat flow in the Poiseuille regime. However, save for two experimental studies of $^4$He crystals by Mezhov-Deglin \cite{mezhov1966,mezhov1967} and by Hogan \textit{et al.} \cite{Hogan1969}, they do not approach the spectacular predictions for size dependence ($\propto d^2$) and for temperature dependence ($\propto T^8$). The emerging picture is the following: in a growing list of solids, one can detect a regime below the $\kappa$ peak where the temperature dependence is faster for thermal conductivity than for specific heat. In other words, with $C \propto T^\gamma$, $\kappa \propto T^{\gamma'}$, one finds $\gamma < \gamma'$ in a narrow temperature window. The correction to the overall amplitude is modest (often between 10 to 30 percent). It is, nevertheless, qualitatively important, because it implies that collisions can help the fluid flow instead of degrading it, as often assumed, when momentum-conserving collisions are neglected. 

Let us briefly review the results. 

\textbf{Solid $^4$He-} At ambient pressure, both isotopes of helium remain a liquid down to $T=0$. Under a pressure of 25 bars (i.e. 2.5 MPa), $^4$He becomes a solid \cite{pobell1996matter}.  Cryogenic crystals can be made under pressures exceeding this pressure. These are remarkable electrical insulators with virtually no chemical impurities and a high degree of isotopic purity. The first confirmation of Gurzhi's theory was reported by Mezhov-Deglin, working at the Institute for Physics Problems and the Institute of Solid State Physics in Moscow in 1966. He measured the thermal conductivity of crystalline $^4$He in the temperature range of 0.5K to 2.5K and at various pressures. In the samples grown at 85 bars, the thermal conductivity was found to peak at 0.9 K and below this peak, it was found to decrease much faster than $T^3$  \cite{mezhov1966}.  A clear peak in the temperature dependence of the effective mean free path could be detected (see Fig.\ref{Fig-old}a). This study was followed by another work by Mezhov-Deglin quantifying the size dependence \cite{mezhov1967} and two other investigations in Duke \cite{Hogan1969} and in Urbana \cite{Hogan1969,thomlinson1969}. The studies led to the confirmation of Poiseuille flow in $^4$He, albeit with a nuance. 

Crystals grown at low pressures (below 100 bars) displayed strong signatures of Poiseuille flow, including a temperature dependence as fast as $\propto T^6$ and a quadratic $\propto d^2$ size dependence \cite{Hogan1969,thomlinson1969}.  The signatures for Poiseuille flow in crystals grown at pressures higher than 185 bars were  weaker \cite{Seward1969}. Nevertheless, these crystals did show a local peak in the temperature dependence of the effective mean free path. Interestingly, the latter crystals did show a prominent Ziman regime and no evidence of any deficit in purity. This suggests to us that phonon hydrodynamics is linked with the proximity of a structural instability, as indicated by posterior research on column V elements. In this line of interpretation, with increasing pressure, the solid state of helium is further stabilized, and therefore Normal phonon-phonon collisions become less frequent, gradually closing the hydrodynamic window.

\textbf{Solid $^3$He-} $^3$He becomes a solid when pressure exceed 30 bars \cite{pobell1996matter}. It crystallizes at low pressures in a Body Centered Cubic (BCC) crystalline phase and then becomes Face Centered Cubic (FCC). In 1969, Thomlinson \cite{thomlinson1969}, almost simultaneously with those studying $^4$He crystals, grew single crystals of $^3$He under pressure and measured their thermal conductivity in both crystalline phases. As seen in Fig.\ref{Fig-old}b, his results led to the observation of a local peak in the temperature dependence of the effective mean free path. 

\textbf{Bismuth-} Crystalline bulk bismuth has fascinating electronic properties. It is a semimetal with an equal density of electrons and holes, which at low temperatures corresponds to one carrier of each sign per 10$^5$ atoms \cite{issi1979}. Because of this low concentration of mobile carriers of charge, the low-temperature thermal conductivity of bismuth has been measured by several authors \cite{kopylov1973,PRATT197874,Boxus1981,Behnia2007}. Like in an insulator, it is almost totally dominated by phonons and peaks between 3 and 4 K.

In 1973, Kopylov and Mezhov-Deglin \cite{kopylov1973}  studied thermal transport in massive bismuth crystals of centimetric dimensions between 1.3 and 7 K. They observed a stronger than cubic temperature-dependent thermal conductivity below the maximum and attributed it to ``a hydrodynamic situation in the phonon system evoked by frequent normal collisions between phonons''. They quantified the mean free path of phonons in their crystals, which showed a local peak, as one can see in Fig. \ref{Fig-old}c. They could not detect any superlinear size dependence in $\kappa$ and in $\ell_{eff}$, expected in the ideal Poiseuille flow.

A few years later, Pratt and Uher \cite{PRATT197874} measured the thermal conductivity of bismuth crystals below 1 K and found that the phonon mean free path shortens with decreasing temperature in agreement with what was found by Kopylov and Mezhov-Deglin \cite{kopylov1973}. They invoked electron-phonon interactions as a source of this. We will return to this feature when discussing the case of antimony. 

\textbf{Solid hydrogen-} Like helium isotopes, solid hydrogen is a quantum crystal, but a molecular one, which at ambient pressure solidifies at $\approx$ 13.8 K. There is a small concentration (0.2\%) of orthohydrogen molecules in equilibrium, which can be
decreased. 

The first study of phonon hydrodynamics in the present century by steady state thermal transport was reported by Zholonko \cite{zholonko2006}, who  measured  the thermal conductivity of solid parahydrogen in the temperature range of 1.5 to 6.0 K. He found that below its low-temperature maximum,  thermal conductivity  displays a rapid variation with an exponent barely larger than 3 and attributed it to a Poiseuille flow of phonons. The estimated mean free path of phonons at $\approx$3 K exceeded the radius of the cylindrical sample and showed a small local peak, barely larger than the noisy background \cite{zholonko2006}. The experimental evidence for Poiseuille flow in this case appears weaker than in other cryogenic crystals.

\begin{figure*}[ht]
\begin{center}
\includegraphics[width=15cm]{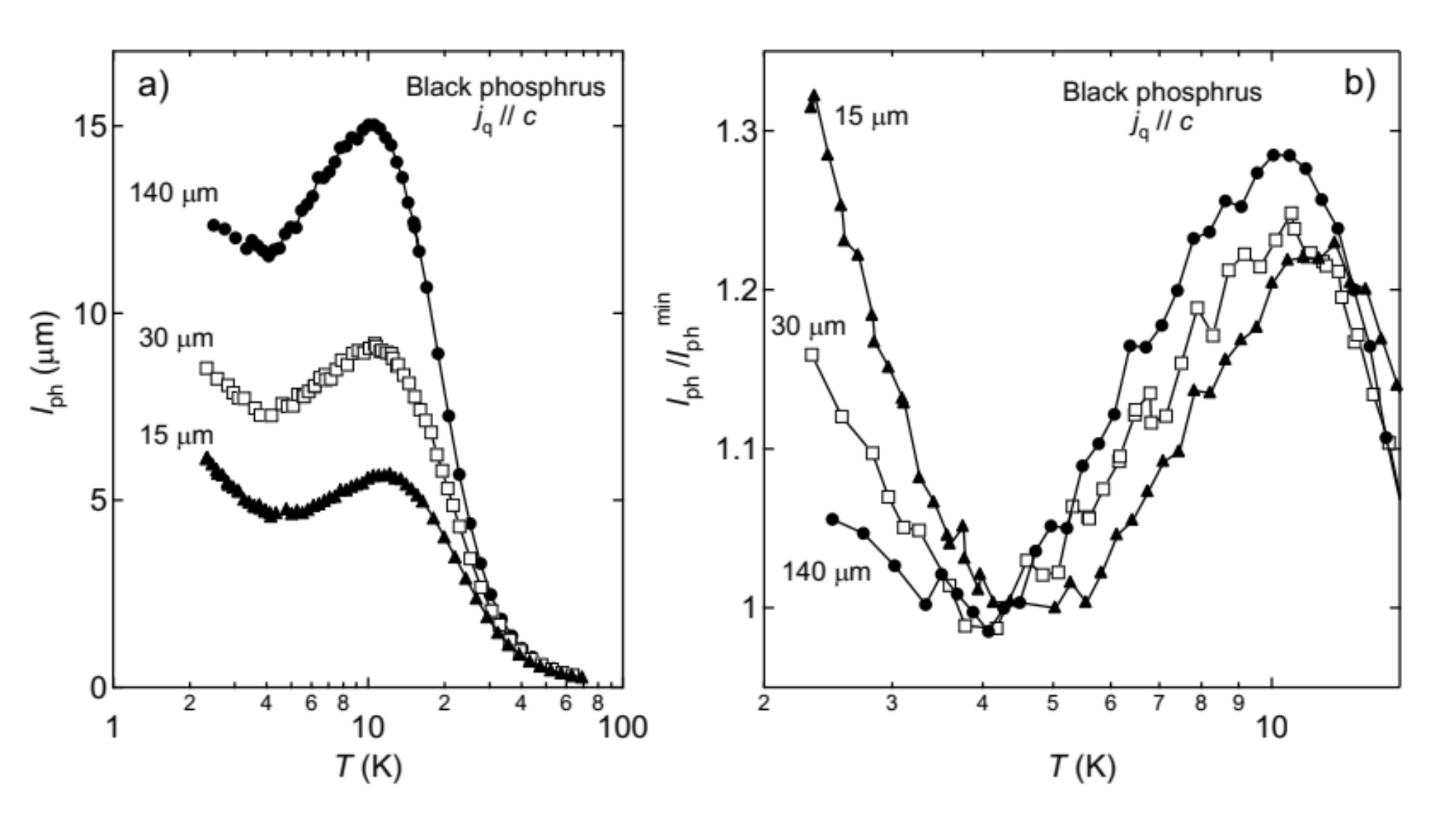}
\caption{\textbf{Poiseuille flow in black P:} a) The extracted phonon mean free path in black P samples with a heat current applied along the  c-axis. The thicker the sample the longer the mean free path.  Note the local peak. b) The same data plotted after normalizing to the minimum mean free path. Note the shift towards lower temperature with increasing thickness, which is expected in the hydrodynamic picture [adapted from ref. \cite{machida2018}].}
\label{Fig-bP}
\end{center}
\end{figure*}

\textbf{Black phosphorus-}  Black phosphorus (BP) is a bulk insulator \cite{Ling2015}, which crystallizes in an orthorhombic crystal structure with puckered honeycomb layers coupled by van der Waals interaction  between the layers. It has a band gap of 2 eV. Like other column V elements and IV-VI binary salts, its crystal symmetry is the outcome of a close competition between between cubic, rhombohedric and orthorhombic  structures\cite{Littlewood_1980,Burdett1981,Behni2016}. Such a proximity to a structural instability may favor Normal phonon-phonon scattering.  

In 2018, Machida and co-workers \cite{machida2018} reported on measurements of thermal conductivity in bulk black phosphorus between 0.1 and 80 K. Several samples with a variety of thicknesses were studied. The study found a thermal conductivity showing a faster than cubic temperature dependence between 5 K and 12 K. The effective phonon mean free path was found to be non-monotonous in temperature dependence in the Poiseuille regime. A local maximum and a local minimum are clearly visible in all samples (See Fig.\ref{Fig-bP}).

The size dependence of the thermal conductivity, in both ballistic and Poiseuille regimes, was found to be close to linear, as previously found in bismuth \cite{kopylov1973}. This is not what one expects in the ideal Poiseuille regime. Nevertheless, a subtle size dependence beyond the ballistic picture was also revealed. It was found that local maximum and the local minimum temperatures both shift with thickness. In  the thicker samples, both occur at lower temperatures. This is understandable in the hydrodynamic picture, because the fraction of phonons suffering frequent Normal collisions is larger in thicker samples (See Fig.\ref{Fig-bP}b).  

The  absence of superlinear size dependence may indicate that the phonon fluid is non-Newtonian \cite{machida2018}. This is because viscosity is set by the Normal scattering rate and this rate is not the same for all phonons. At a given temperature, Normal scattering is much less frequent for phonons close to the sample boundary. It has been shown that the velocity profile of  a non-Newtonian fluid is significantly flatter than the parabolic one, which is expected for Newtonian fluids \cite{Eu1990}. This would explain a size dependence less straightforward and drastic than what is expected in the simple and ideal Poiseuille flow. 

\textbf{Strontium titanate-} SrTiO$_3$, known sometimes simply as STO, this  large-gap insulator with a perovskite structure,  is  a quantum paraelectric, where zero-point quantum fluctuations play a crucial role in the stabilization of the ground state \cite{Muller1979}. An extremely small concentration of chemical dopants turn this insulator to a metal with a superconducting instability \cite{collignon2019}.  Its thermal conductivity was first measured by Steigmeier \cite{Steigmeier1968}, who discovered that electrical field can modify the amplitude of thermal conductivity at cryogenic temperatures. 

In 2018, Martelli \textit{et al.} \cite{martelli2018} reported on a study of thermal conductivity in SrTi$_{1-x}$Nb$_x$O$_3$.  All four regimes of heat transport could be detected in undoped STO. Moreover, it was found that three different samples showed a $\kappa$ evolving faster than cubic with temperature below its peak and in a narrow temperature window. The behavior was attributed to a Poiseuille flow of phonons. It was found to disappear in the presence of Nb dopants. 

\begin{figure*}[ht]
\begin{center}
\includegraphics[width=15cm]{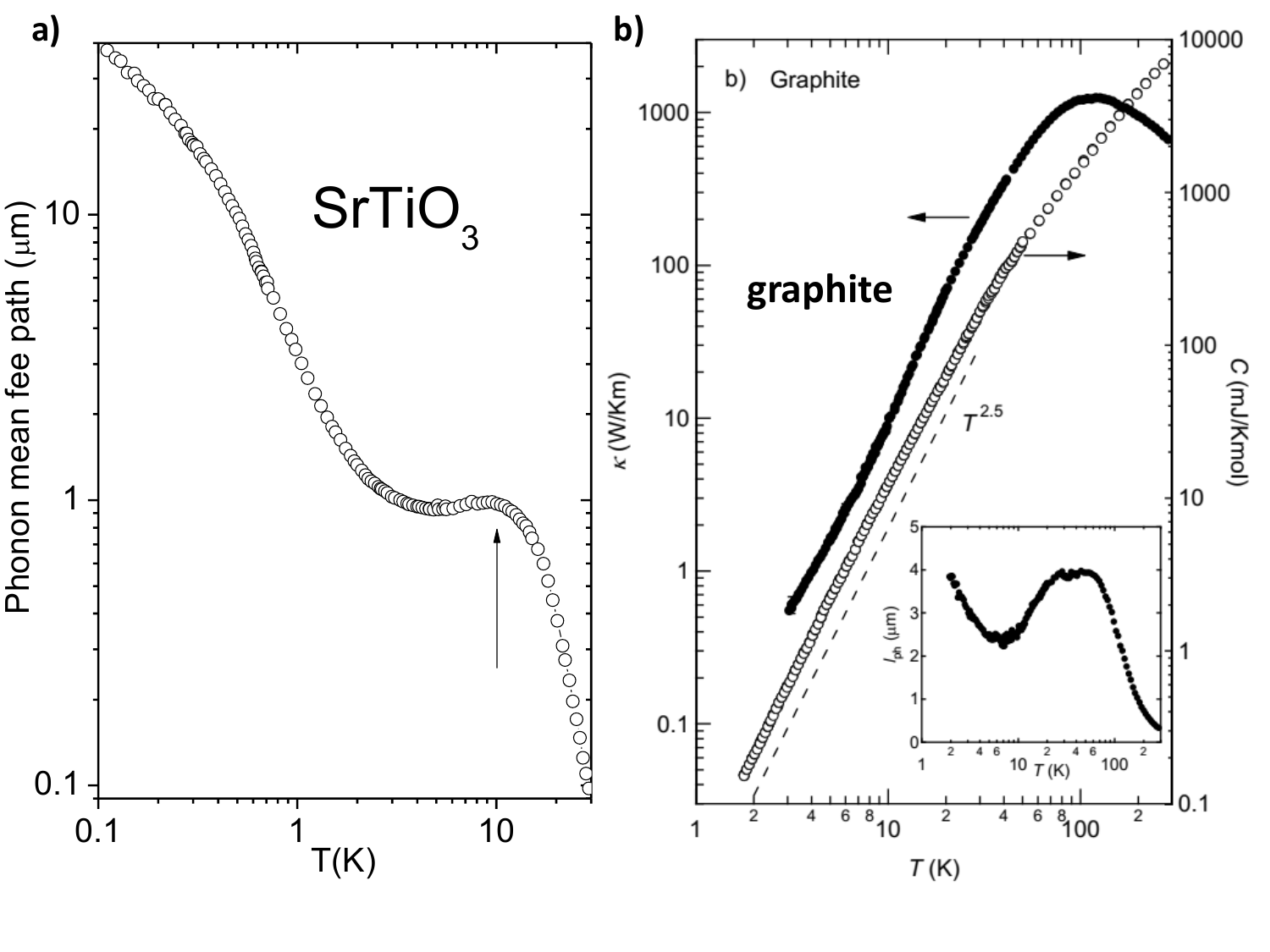}
\caption{\textbf{Two cases of non-monotonous phonon mean free path-} a) Temperature dependence of the mean free path of phonons in strontium titanate, estimated by dividing the thermal conductivity by  product of the specific heat and the sound velocity ($\ell_{ph}=\frac{3\kappa}{v_s C}$) \cite{martelli2018,Jaoui2023}. The mean free path remains well below the thickness of the sample. It shows a local maximum shown by an arrow. b) Temperature dependence of the in-plane thermal conductivity and the specific heat in graphite. Thermal conductivity evolves faster than the specific heat. As seen in the inset, the mean free path of phonons displays a peak \cite{machida2020}.}
\label{Fig-STO-HOPG}
\end{center} 
\end{figure*}

It is noteworthy  that in strontium titanate, at T$\approx$ 2K, the thermal conductivity is almost cubic, but  the phonon mean free path is still orders of magnitude shorter than the crystal dimensions in the Poiseuille regime. It is of the order of a few microns, the estimated size of the tetragonal structural domains \cite{Buckley1999}. Therefore, the ballistic regime is not attained above 0.1 K. Nevertheless, as seen in Fig. \ref{Fig-STO-HOPG}a,  the phonon mean free of a typical undoped STO crystal displays a clearly non-monotonous temperature dependence. 

The peculiarity of SrTiO$_3$ is the presence of two soft phonon modes \cite{Yamada1969}. The one at the center of the Brillouin zone is associated with ferroelectric instability and its frequency does not vanish at zero temperature. It has a significant dispersion and contributes to heat transport. The other at the zone boundary drives  the structural transition at 105 K. A recent inelastic neutron scattering study \cite{Fauque2022} has found evidence for hybridization between a transverse acoustic mode and the zone center ferroelectric soft mode. This may play a role in the emergence of phonon hydrodynamic phenomena. 

\textbf{Graphite-} Allotropes of carbon are among the best known conductors of heat. At room temperature, the thermal conductivity of diamond is superior to that of copper \cite{pobell1996matter}. The other carbon allotrope, graphite,  has a layered structure combining strong interlayer covalent with weak intralayer van der Waals bonds. The two-hundred-fold difference in the strength of bonding leads to two distinct in-plane and out-of-plane Debye energy scales. This leads to a specific heat which in graphite, evolves slower than $T^3$ \cite{Khrumhansl1953}. This has been known for a long time and discussed as an exception illuminating the Debye $T^3$ rule in a textbook like Ziman's \cite{ziman2001electrons}. 

In 2020, decades after the previous studies of heat transport in graphite \cite{Slack1962,HOLLAND1966903,Taylor1966-TAYTTC-4}, Machida \textit{et al.} \cite{machida2020} carried out a study of thermal transport in thin graphite samples  and found that the in-plane thermal conductivity varies faster than the specific heat (See Fig.\ref{Fig-STO-HOPG} b). This was unnoticed in previous studies. Moreover, the effect was found to become more pronounced with thinning of the graphite samples, approaching what was reported in the case of graphene \cite{Balandin2008}.

The origin of the pronounced size dependence of the thermal conductivity in thin graphite is still to be properly understood. It is a subject of ongoing theoretical research \cite{Guo2023}.

\begin{figure*}[ht]
\begin{center}
\includegraphics[width=16cm]{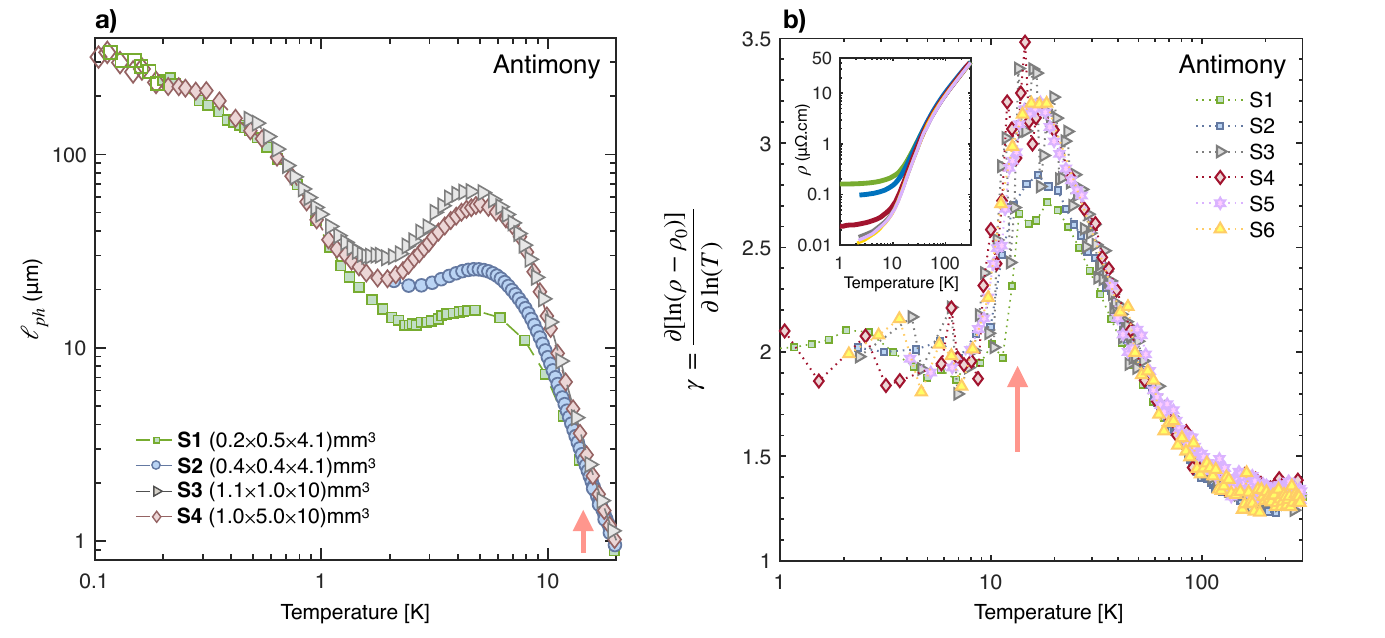}
\caption{\textbf{The case of antimony-} a) The temperature dependence of the mean free path $\ell_{ph}$ in single crystals of antimony with different dimensions. The heat current is applied along the bisectrix axis. There is a local maximum, which is amplified in larger samples. The arrow highlights the temperature at which the phonon mean free path becomes size-dependent. Below this temperature, Umklapp events are rare and ph-ph collisions conserve momentum. Note that below 1 K, the phonon mean free path becomes sample independent again. b) The temperature dependence of the exponent, $\gamma$, of electrical resistivity in different samples (shown in the inset). Despite the difference in the magnitude of resistivity at low temperature, $\gamma$ (extracted from $\rho=\rho_0+T^\gamma$) shows a similar behavior in all of them. With decreasing temperature, it steadily increases from 1 towards 5, as expected in the Bloch-Gr\"uneisen picture. But this steady increase in abruptly stopped when ph-ph collisions cease to be resistive. The arrows in the two panels point to the same temperature (adapted from ref. \cite{Jaoui2022}). }
\label{Fig-Sb}
\end{center}
\end{figure*}

\section{Phonon-electron hydrodynamics}
The phonon and electron fluids can exchange momentum through Normal collisions between these two types of quasi-particles. As a result, hydrodynamic effects involving both fluids can emerge. One known example is phonon drag, a non-diffusive thermoelectric response generated by heat-carrying phonons coupled to the electron bath, which dominates the low temperature thermoelectric response of many semiconductors \cite{Herring1954}.

Evidence for frequent phonon-electron momentum exchange with signatures in thermal and electrical transport channels was uncovered by a study of electrical and thermal transport in antimony (Sb) crystals of various sizes~\cite{Jaoui2022}. Jaoui and co-workers found that the phonon thermal conductivity does not show any size dependence, in contrast to the electronic thermal and electrical conductivities. How come that the mean free path of electrons increases with the sample size, but not the mean free path of phonons? In other words, if disorder is at work, why does it affects phonons but not electrons? It was also found that phonon thermal conductivity displays prominent quantum oscillations, implying that the evacuation of each Landau level affects the phonon mean free path. Taken together, these facts point to a case of frequent momentum-conserving collisions between electrons and phonons at cryogenic temperatures. Coupled electron-phonon hydrodynamics has been a subject of emerging theoretical interest \cite{Coulter2019,levchenko2020,Huang2021}. 

Fig.\ref{Fig-Sb}a shows the temperature dependence of the phonon mean free path in several crystals of antimony of various sizes. It shows a local maximum, like in other cases of phonon Poiseuille flow. There is a clear size dependence near this peak, but in addition to a high temperature intrinsic regime, there is a second one at low temperature with no detectable size dependence in the phonon mean free path. This is where the scattering of phonons by electrons dominates the lattice heat transport.

It is interesting to look at the situation from the point of view of electrons. At low temperatures, the electrical resistivity, as well as the electronic thermal resistivity of antimony crystals, is dominated by electron-electron scattering \cite{Jaoui2021}. This leads to quadratic temperature dependence of both, with a prefactor, which evolves with the sample size \cite{Jaoui2021}. As a result, the Lorenz number becomes size-dependent, as theoretically expected in the case of electronic hydrodynamics \cite{principi2015}. 

The strong coupling between electrons and phonons leads to an abrupt breakdown of the Bloch-Grüneisen picture \cite{ziman2001electrons}  below 15 K. This is shown in Fig.\ref{Fig-Sb}b, which plots the temperature dependence of the exponent of the electrical resistivity ($\gamma$, with $\rho=\rho_0+AT^\gamma$). In this picture, $\gamma$ is expected to smoothly evolve from $\gamma\simeq 1$ above the Debye temperature to  $\gamma\simeq 5$ at temperatures much below. In the high temperature regime, all phonons are excited and electron-phonon scattering is essentially elastic.  At low temperatures, the phonon population diminishes and the typical phonon wave-vector shrinks, pulling down the e-phonon cross section. This is indeed what happens in noble metals, copper, silver or gold. In contrast, as one can see in Fig.\ref{Fig-Sb}b, in antimony, the steady increase in $\gamma$ is suddenly interrupted in antimony when phonons do not suffer Umklapp collisions and scattering by phonons ceases to decay the flow of electrons. Thus, there is an asymmetry between the two components of the electron-phonon bifluid. Momentum can leak from the electron bath, but not from the phonon bath. Whenever an electron yields momentum to the phonon fluid, the electron fluid is going to receive it back in another momentum exchange event. The only source of inelastic dissipation is scattering between electrons. 

\section{Open questions}
In the previous century, it was believed that the observation of phonon Poiseuille flow required single crystals of exceptional purity. The findings of the past few years indicate that this is not necessarily the case.  When Normal collisions are frequent enough, then even in the presence of disorder,  phonon hydrodynamics becomes detectable.  Two remarkable cases are strontium titanate and graphite. These two solids are far from being isotopically pure. Moreover, and for different reasons, even in cryogenic temperatures, their phonon mean free path remains well below the sample size. Therefore, their ballistic regime is elusive. Nevertheless, not only one can detect signatures of the Poiseuille flow, as we saw above, but also manifestations of the second sound have been experimentally observed by optical probes in both 
materials \cite{Huberman2019,Ding2022,koreeda2023coherent}. 

But what sets the amplitude of Normal collisions? Early works on graphene \cite{Cepellotti2015,Lee2015} highlighted the quadratic dispersion of acoustic phonons, which amplify the phonon density of states. More recently, hybridization between acoustic and optical phonons have been observed in strontium titanium \cite{Fauque2022} and may play a role in the amplification of momentum exchange between phonons. In the case of column V elements, Bi, Sb and black P, one suspects the proximity of a structural instability \cite{Littlewood_1980,Behni2016} to be at work.  One can see that the suspicions go along different directions and one may wonder if a common ground exists between them.

Another open question is the possible link between phonon hydrodynamics and the phonon thermal Hall effect. The reason why phonons in some insulators generate a detectable thermal Hall response is a mystery. Is it a coincidence that the two non-magnetic insulators, which display a thermal Hall effect are strontium titanate \cite{li2020} and black P \cite{li2023}, which were both found to host Poiseuille flow of phonons? Future studies will tell.

One may also speculate on a link between this thematic and the recent theoretical progress \cite{Simoncelli2019,Isaeva2019} in unifying the pictures of heat transport in crystals and glasses, which  considering the wave nature of phonons and diffusion across phonon branches. 

\section{Acknowledgements}
This work was supported in France by JEIP-Coll\`{e}ge de France, by the Agence Nationale de la Recherche (ANR-19-CE30-0014-04), a grant by the Ile de France regional council, and Fundação de Amparo à Pesquisa do Estado de São Paulo (2018/19420-3).
\bibliography{ref}
\end{document}